\documentclass[apjl]{emulateapj}
\usepackage{amsfonts,amsmath,graphicx,natbib,apjfonts,color,gensymb,longtable}

\def\swift{{\it Swift}}
\def\sax{{\it BeppoSAX}}

\def\fe{1}
\def\icranet{2}
\def\bo{3}
\def\nw{4}

\shorttitle{FRB\,010724 and BeppoSAX}
\shortauthors{Guidorzi~et~al.}

\begin{document}
\title{A search for gamma--ray prompt emission associated with the Lorimer Burst FRB\,010724}

\author{C.~Guidorzi\altaffilmark{\fe}, M.~Marongiu\altaffilmark{\fe,\icranet}, R.~Martone\altaffilmark{\fe,\icranet}, L.~Amati\altaffilmark{\bo}, F.~Frontera\altaffilmark{\fe,\bo}, L.~Nicastro\altaffilmark{\bo}, M.~Orlandini\altaffilmark{\bo}, R.~Margutti\altaffilmark{\nw}, E.~Virgilli\altaffilmark{\fe}
}
\altaffiltext{\fe}{Department of Physics and Earth Science, University of Ferrara, via Saragat 1, I-44122, Ferrara, Italy}
\altaffiltext{\icranet}{ICRANet, Piazza della Repubblica 10, I-65122 Pescara, Italy}
\altaffiltext{\bo}{INAF--Osservatorio di Astrofisica e Scienza dello Spazio di Bologna, Via Piero Gobetti 93/3, I-40129 Bologna, Italy}
\altaffiltext{\nw}{Center for Interdisciplinary Exploration and Research in Astrophysics and Department of Physics and Astronomy, Northwestern University, 2145 Sheridan Road, Evanston, IL 60208-3112, USA}

\begin{abstract}
No transient electromagnetic emission has yet been found in association to fast radio bursts (FRBs), the only possible exception ($3\sigma$ confidence) being the putative $\gamma$--ray signal detected in \swift/BAT data in the energy band 15--150~keV at the time and position of FRB\,131104. Systematic searches for hard X/$\gamma$--ray counterparts to other FRBs ended up with just lower limits on the radio/$\gamma$--ray fluence ratios. In 2001, at the time of the earliest discovered FRBs, the {\it BeppoSAX} Gamma-Ray Burst Monitor (GRBM) was one of the most sensitive open sky $\gamma$--ray monitors in the 40--700~keV energy band. During its lifetime, one of the FRBs with the highest radio fluence ever recorded, FRB\,010724 ($800\pm400$~Jy\,ms), also known as the ``Lorimer burst'', was promptly visible to the GRBM. Upon an accurate modeling of the GRBM background, eased by its equatorial orbit, we searched for a possible $\gamma$--ray signal in the first 400~s following the FRB, similar to that claimed for FRB\,131104 and found no significant emission down to a $5\sigma$ limit in the range $(0.24$-- $4.7)\times10^{-6}$~erg~cm$^{-2}$ (corresponding to 1 and 400~s integration time, respectively), in the energy band 40--700~keV. This corresponds to $\eta = F_{\rm radio}/F_{\gamma}>10^{8-9}$~Jy\,ms\,erg$^{-1}$\,cm$^2$, i.e. the deepest limit on the ratio between radio and $\gamma$--ray fluence, which rules out a $\gamma$--ray counterpart similar to that of FRB\,131104. We discuss the implications on the possible mechanisms and progenitors that have been proposed in the literature, also taking into account its relatively low dispersion measure ($375\pm3$\,pc\,cm$^{-3}$) and an inferred redshift limit of $z<0.4$.
\end{abstract}

\section{Introduction}
\label{sec:intro}
Fast radio bursts (FRBs) are ms-long bright pulses of unknown origin characterized by a dispersion measure (DM) significantly in excess of the corresponding Galactic value \citep{Lorimer07,Thornton13}. A large DM value, due to a relatively large electron column density integrated along the sightline, strongly hints at an extragalactic origin. Despite the high all-sky daily rate of several thousands (e.g., \citealt{Rane16,Crawford16,Scholz16}), only $\sim80$~FRBs have been publicly announced so far (as of July 2019; \citealt{Petroff16}). This is mainly due to the relatively narrow fields of view (FOV) of most of the radiotelescopes that first observed them, although the discovery rate is now ramping up thanks to a new generation of wide FOV experiments that have recently come on stage. The majority of them appear to be one-off events, except for a couple of them \citep{Spitler16,CHIME19a}. See  \citet{Petroff19_rev,Katz18rev,Popov18_rev,RaneLorimer17_rev} for updated reviews.

Affected by several arcmin positional uncertainty, FRBs so far defied any search for either prompt simultaneous or afterglow transient emission at other wavelengths that would help identify their host galaxies, measure the distance, constrain the radiation mechanism and the nature of the progenitor. There are a few exceptions: FRB\,121102, whose host was found at a redshift of $z=0.193$ \citep{Tendulkar17}, which is also one of the two repeating sources: it was precisely its repeating nature that made it possible to localize with sub-arcsec accuracy during interferometric radio observations \citep{Chatterjee17}. Recently, the redshifts of two other non-repeating FRBs have been determined: FRB\,180924 at $z=0.3214$ \citep{Bannister19} and FRB\,190523 at $z=0.66$ \citep{Ravi19b}.

A possible ($3\sigma$ confidence) $\gamma$--ray $\sim400$-s long transient source, Swift~J0644.5-5111, positionally and temporally compatible with FRB\,131104 was reported by \citet[hereafter D16]{DeLaunay16} in the 15--150~keV band within the data of the Burst Alert Telescope (BAT; \citealt{Barthelmy05}) aboard the {\em Neil Gehrels Swift Observatory} \citep{Gehrels04}.
The possibility of a gamma--ray burst (GRB) associated to this FRB was later called into question by the lack of any radio afterglow \citep{ShannonRavi17}. However, admitting microphysical shock parameter values different from what is typically assumed for GRB afterglows, this GRB--FRB association remains plausible \citep{GaoZhang17}. A similar search carried out by D16 for other FRBs that were promptly visible with BAT yielded only upper limits. Previous searches for high--energy counterparts to FRBs had ended up with lower limits to the radio/$\gamma$ fluence ratio $\eta_{\rm frb}>10^{7-9}$~Jy\,ms\,\,erg$^{-1}$\,cm$^2$, incompatible with the upper limit $\eta_{\rm sgr}<10^{7}$~Jy\,ms\,\,erg$^{-1}$\,cm$^2$ derived for the $\gamma$--ray giant flare of the Galactic magnetar SGR\,1806-20 \citep{Tendulkar16}.
Similarly, \citet{Scholz17} analyzed the data of the Gamma--ray Burst Monitor \citep{Meegan09} aboard the {\em Fermi Gamma--Ray Telescope} and found nothing down to $10^{-8}$\,erg\,cm$^{-2}$ in the 10--100~keV energy band in a 2--s window centered on the arrival times of four bursts of repeater FRB\,121102.
Reversing the approach, \citet{Palaniswamy14} found no prompt radio pulses compatible with FRBs associated to five GRBs that were observed starting within 140~s of the $\gamma$--ray trigger time.
No detections were reported from searches in the GeV domain using the {\em Fermi} Large Area Telescope \citep{Atwood09} for both one-off FRBs \citep{Yamasaki16,Xi17} and repeaters \citep{ZhangZhang17,Yang19}.
Analogous results were obtained at very high energies, above 100~GeV \citep{HESS17,MAGIC18} as well as for TeV--PeV neutrinos coincident with FRBs \citep{IceCube18_FRB,ANTARES19}.

One of the open questions is whether a common family of progenitors is responsible for the observed population of FRBs, in particular one-off and repeating sources \citep{Palaniswamy18,Caleb19}, or, as in the case of GRBs, at least two families are required to explain the long and short duration events respectively associated to the core collapse of some kind of hydrogen-stripped massive stars and to the merging of compact objects (see \citealt{KumarZhang15rev} for a review). Thus, the existence of repeating sources does not necessarily rule out the so-called cataclysmic models.
A number of the proposed models that predict an associated high-energy emission invoke rapidly rotating young neutron stars (NS) and magnetars as progenitors \citep{Katz14,CordesWassermann16,Lyutikov16a}.
Given also that in some models ms magnetars are supposed to form following the core-collapse that powers both long \citep{Usov92,Thompson94,Bucciantini07,Metzger11} and short GRBs \citep{FanXu06,Metzger08,Rowlinson10,Gompertz13}, as well as superluminous SNe (SLSNe; \citealt{Nicholl17,Metzger17}), it is worth searching for hard X/$\gamma$--ray emission associated to FRBs. As with GRBs and SNe the study of the environments may potentially provide useful clues on the homogeneity of the FRB progenitors \citep{EftekhariBerger17}.

Motivated by the results of D16, we exploited the sensitivity of the Gamma--Ray Burst Monitor (GRBM; \citealt{Frontera97}) that operated as an open--sky detector in the energy band $40$--$700$~keV aboard {\em BeppoSAX} \citep{Boella97} in 1996--2002. We carried out a detailed search for a high-energy counterpart to FRB\,010724, the first reported FRB \citep{Lorimer07}, popularly known as the Lorimer burst: this was so bright that it saturated the Parkes multibeam receiver and was also detected in three other beams of the receiver. From an initial estimate of $150$\,Jy\,ms, its fluence has recently been revised to $800\pm400$\,Jy\,ms \citep{Ravi19}. Owing to its exceptional brightness, FRB\,010724 might represent a discovery bias \citep{MacquartEkers18a} and may come from a space-limited population, whose local density would far exceed the cosmological one \citep{Katz16b}. Given its relatively low DM of $375\pm3$\,pc\,cm$^{-3}$, evidence that this FRB might be relatively nearby is also provided by the brightness-dispersion correlation shown in the ASKAP sample (\citealt{Shannon18}; see also \citealt{Niino18}).
In this respect, another example of low-DM/high-flux FRB is given by FRB\,110214 \citep{Petroff19}. We therefore discuss the implications of our results on the proposed associations as a function of the possible redshift compatibly with its DM. Hereafter, we assume the latest Planck cosmological parameters: $H_0=67.74$~km\,s$^{-1}$\,Mpc$^{-1}$, $\Omega_m=0.31$, $\Omega_\Lambda=0.69$ \citep{cosmoPlanck15}.

\section{Data analysis}
\label{sec:data}
From the FRB catalog\footnote{\url{http://www.frbcat.org}} \citep{Petroff16} we selected the four FRBs occurred in 2001 during the \sax\ operational life (Table~\ref{tab:frbsax}) and checked the availability of GRBM data at the time of each FRB along with its visibility (i.e., not Earth-blocked). Only FRB\,010724 passed the selection.
\begin{deluxetable*}{lcrrrrrcl}
\tablecolumns{3}
\tablewidth{0pc}
\tablecaption{\sax/GRBM data availability for the FRBs detected in 2001. \label{tab:frbsax}}
\tablehead{
\colhead{FRB} & \colhead{UT} & \colhead{RA(J2000)} & \colhead{DEC(J2000)} & \colhead{$\phi$} & \colhead{$\theta$} & \colhead{Elevation} & \colhead{OP} & \colhead{Comment}\\
\colhead{} & \colhead{} & \colhead{($^\circ$)} & \colhead{($^\circ$)} & \colhead{($^\circ$)} & \colhead{($^\circ$)} & \colhead{($^\circ$)} & \colhead{} & \colhead{}
}
\startdata
010125 & 00:29:14 & $286.7$ &  $-40.7$  &  $42.2$  &  $19.4$  &  $-18.0$ & 10600 & Earth blocked\\
010312\tablenotemark{a} & 11:06:48 & $ 81.7$ &  $-64.9$  &   --     &    --    &  $  1.4$ & 10862 & no data\\
010621 & 13:02:09 & $283.0$ &  $ -8.5$  & $203.5$  &  $-8.8$  &  $ 59.5$ & 11411 & unavailable due to data gap\\
010724 & 19:50:00 & $ 19.5$ &  $-75.2$  & $162.0$  & $-43.5$  &  $ 38.6$ & 11651 & visible and available
\enddata
\tablenotetext{a}{Reported by \citet{Zhang19a}.}
\tablecomments{($\phi$, $\theta$) is the direction with respect to the \sax\ payload reference frame \citep{Frontera09}. The elevation angle is calculated with respect to the Earth limb. OP is the observing period, which identifies the data set in the \sax\ archive.}
\end{deluxetable*}
Based on the direction of FRB\,010724 with reference to the \sax\ frame, the expected signal for each of the four GRBM units is comparable with one another (Table~\ref{tab:expected}). From the \sax\ GRB catalog \citep{Frontera09} we used the response matrices of the GRB with the nearest local position $(158^\circ, -46^\circ)$ to that of the FRB $(162^\circ, -43.5^\circ)$, and for each unit we calculated the expected number of counts for both energy bands $40$--$700$~keV and $>100$~keV, assuming the same spectral models and $\gamma$--ray fluences found by D16 for FRB\,131104 (Table~\ref{tab:expected}). No prompt \sax\ X--ray data are available for FRB\,010724, since it lies outside the FOV of the Wide Field Cameras \citep{Jager97}.

We modeled the background rates for each unit and energy band by exploiting the GRBM data sharing the most similar configuration and that were  acquired about one day before the FRB. A detailed description is reported in Appendix~\ref{sec:bkg}.
\begin{deluxetable*}{lcrrrrrrrrr}
\tablecolumns{11}
\tablewidth{0pc}
\tablecaption{Expected net counts for a $\gamma$--ray event equal to that found for FRB\,131104\tablenotemark{a}.
\label{tab:expected}}
\tablehead{
\colhead{Model\tablenotemark{b}} & \colhead{$\Gamma$} & \colhead{$kT$} & \colhead{$F_\gamma(40-700)$} & \colhead{Band} & \colhead{GRBM1} & \colhead{GRBM2} & \colhead{GRBM3} & \colhead{GRBM4} & \colhead{Mean} & \colhead{$5\sigma$ U.L. on $F_\gamma$\tablenotemark{c}}\\
\colhead{} & \colhead{} & \colhead{ (keV)} & \colhead{(erg\,cm$^{-2}$)} & \colhead{(keV)} & \colhead{(counts)} & \colhead{(counts)} & \colhead{(counts)} & \colhead{(counts)} & \colhead{(counts)} & \colhead{(erg\,cm$^{-2}$)}
}
\startdata
{\sc pow} & $1.16$\tablenotemark{d} &  --  & $1.5\times10^{-5}$ & 40--700 &  4698 &  3573 &  4916 &  7248 &   5109 & $<4.0\times10^{-6}$ \\
          &        &  --  &                    & $>$ 100 &  4286 &  3129 &  4381 &  7032 &   4707 & \\
{\sc pow} & $0.38$\tablenotemark{d} &  --  & $4.9\times10^{-5}$ & 40--700 & 16754 & 11971 & 17188 & 25820 &  17933 & $<3.7\times10^{-6}$\\
          &        &  --  &                    & $>$ 100 & 16303 & 11370 & 16722 & 26938 &  17833 & \\
{\sc pow} & $1.84$\tablenotemark{d} &  --  & $6.1\times10^{-6}$ & 40--700 &  1603 &  1316 &  1696 &  2392 &   1752  & $<4.7\times10^{-6}$\\
          &        &  --  &                    & $>$ 100 &  1304 &  1024 &  1338 &  2122 &   1447 & \\
{\sc pow} & $2$    &  --  & $5.0\times10^{-6}$ & 40--700 &  1259 &  1055 &  1332 &  1852 &   1374  & $<5.0\times10^{-6}$\\
          &        &  --  &                    & $>$ 100 &   991 &   793 &  1012 &  1600 &   1099 & \\
{\sc tb}  &  --    & $200$& $4.2\times10^{-6}$ & 40--700 &  1017 &   896 &  1097 &  1472 &   1120  & $<4.7\times10^{-6}$\\
          &  --    &      &                    & $>$ 100 &   741 &   628 &   760 &  1182 &    828 & \\
{\sc tb}  &  --    &  $75$& $2.3\times10^{-6}$ & 40--700 &   410 &   405 &   439 &   522 &    444  & $<7.1\times10^{-6}$\\
          &  --    &      &                    & $>$ 100 &   227 &   218 &   219 &   325 &    247 &
\enddata
\tablenotetext{a}{Counts are meant to be in excess of the background for each GRBM unit for a $\gamma$--ray event with $F_\gamma(15-150\ \rm{keV})=4\times10^{-6}$~erg~cm$^{-2}$.}
\tablenotetext{b}{The spectral models are the same as in D16: power--law ({\sc pow}) and thermal bremsstrahlung ({\sc tb}).}
\tablenotetext{c}{Upper limits obtained by integrating the background-subtracted mean counts over the $[0,400]$~s interval.}
\tablenotetext{d}{Photon index values corresponding to the confidence interval estimated by D16.}
\end{deluxetable*}
%

\section{Results}
\label{SubSec:results}
\begin{figure}
\begin{center}
\scalebox{1.}
{\includegraphics[width=0.5\textwidth]{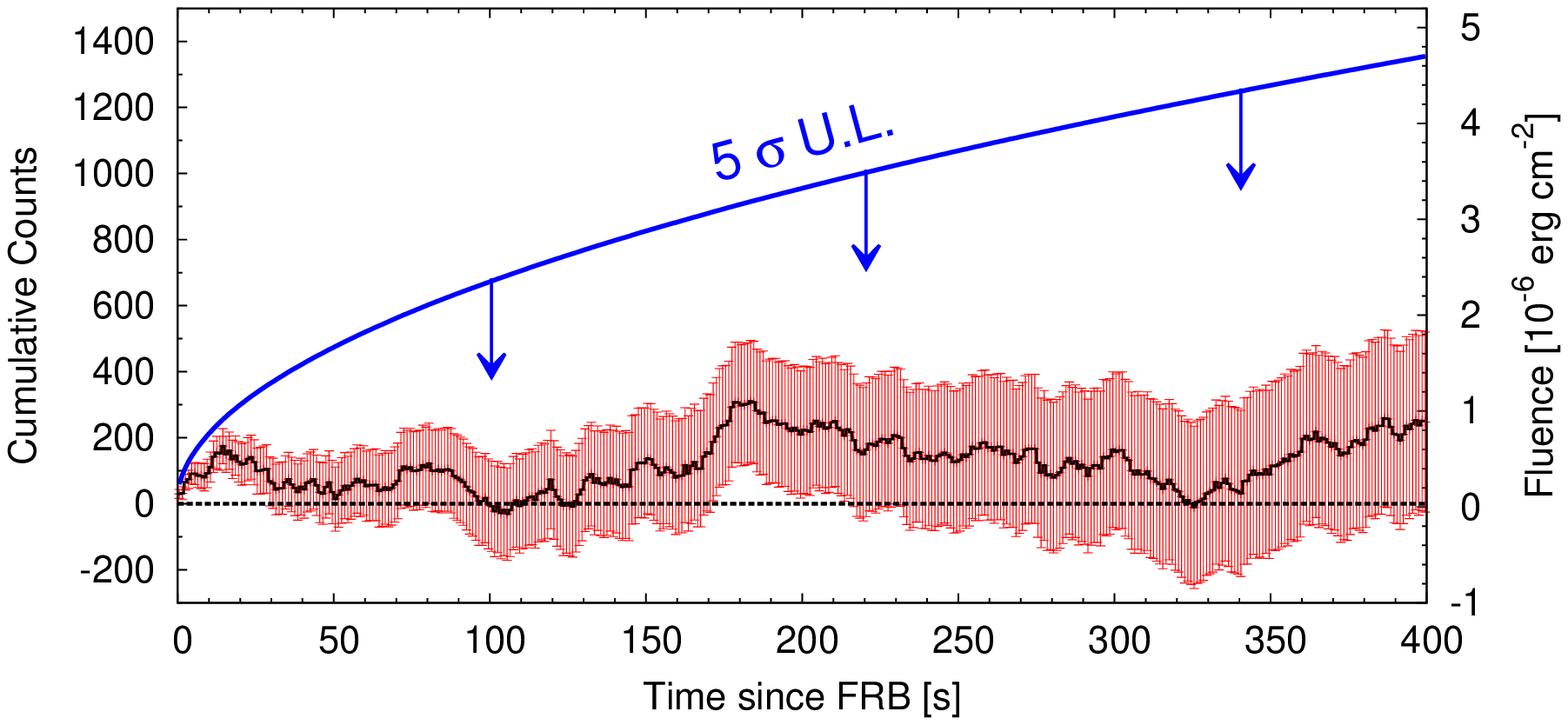}}
\caption{Cumulative net counts in the 40--700~keV energy band averaged over the four GRBM units as a function of integration time, from 0 to $t$, with $t$ spanning the 0-400~s interval. Error bars are $1\sigma$. The solid blue line shows the corresponding $5\sigma$ upper limits. The corresponding fluence (right-hand axis) is calculated assuming a power--law spectrum with $\Gamma=1.84$.}
\label{fig:cumfluence}
\end{center}
\end{figure}

In order to investigate a potential $\gamma$--ray counterpart, we assumed a fast-rise slow-decay pulse profile as modeled by \citet{Kocevski03} so as to have the desired duration of $\sim400$~s. We chose the following parameters: rise and decay indices $r=2$ and $d=3$, respectively; peak time $t_m=100$~s and $t_0=0$. Normalizations were chosen so as to match the expected counts (Sect.~\ref{sec:data}) in the time interval from 0 to 400~s, comparable with the $T_{90}=377$~s of the \swift\, event. The resulting profiles are shown in %
\begin{figure}
\begin{center}
\scalebox{1.0}
{\includegraphics[width=0.5\textwidth]{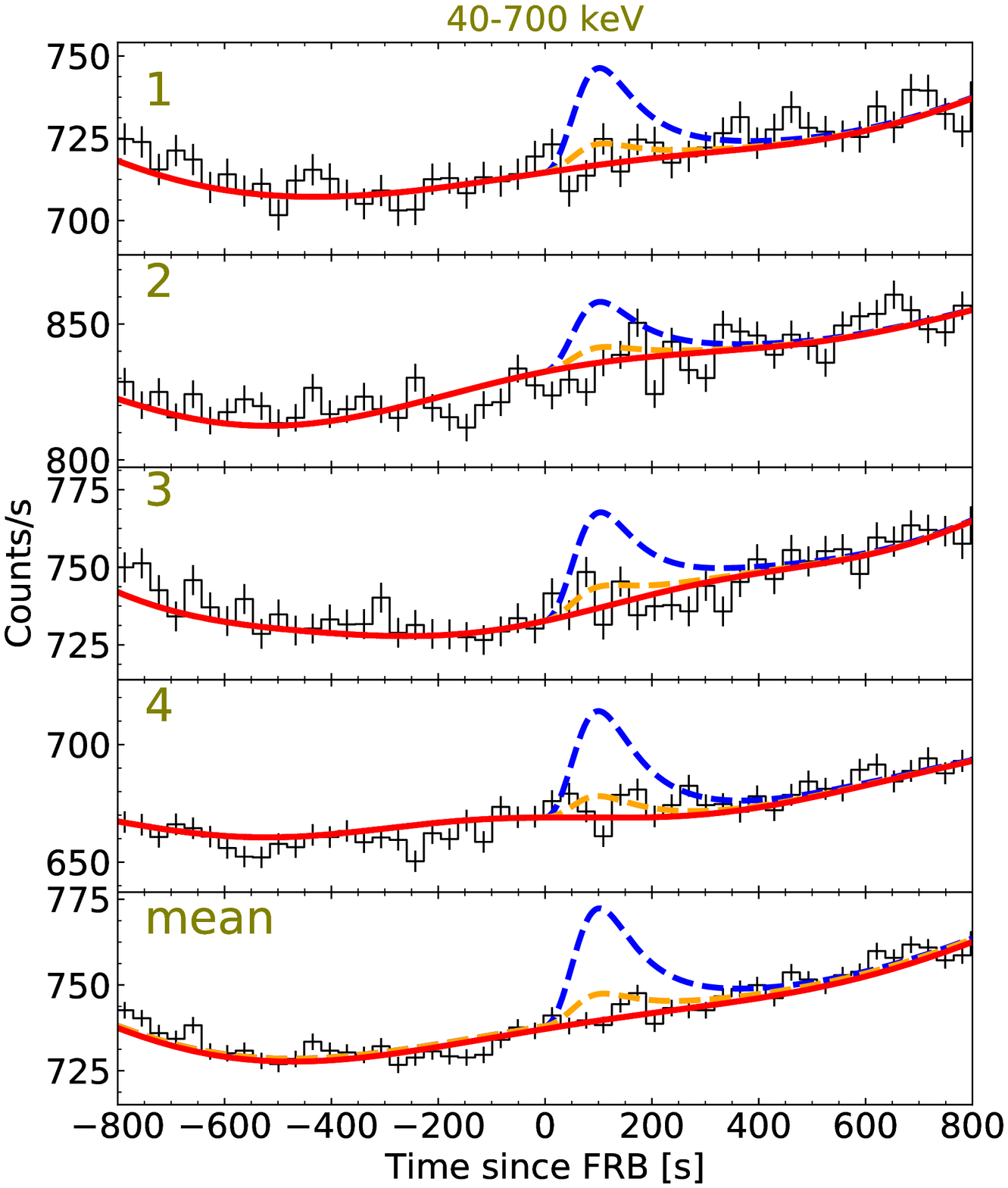}}
\caption{\sax/GRBM ratemeters (40--700~keV; bin time of 32~s) at the time of FRB\,010724. Background (solid lines) was obtained modeling a different orbit. From top to bottom: individual units 1 to 4, the mean of the four units. Dashed blue (orange) profile is what one would expect for an event equal to that found by D16 in \swift/BAT data for FRB\,131104 extrapolating the best-fit power--law (thermal bremsstrahlung) model to the 40-700 keV band.}
\label{fig:all32s}
\end{center}
\end{figure}
Figure~\ref{fig:all32s} with dashed lines, added to the background level, for the 40--700~keV channels of each of the detectors, as well as for the average of the four, which has the best signal-to-noise ratio (S/N). Specifically, blue and orange lines correspond to the best-fit power--law (hereafter {\sc pow}) and thermal bremsstrahlung (hereafter {\sc tb}) models, respectively.
Furthermore, we integrated the background-subtracted counts in the 0--400~s time intervals and found no significant ($>5\sigma$) excess. The most constraining upper limit to the 40--700~keV fluence is obtained from the mean time profile: a total count fluence of $247\pm273$ turns into a $5\sigma$ upper limit of $F_\gamma(40-700)<4.7\times10^{-6}$~erg~cm$^{-2}$ for the {\sc pow} and $F_\gamma(40-700)<7.1\times10^{-6}$~erg~cm$^{-2}$ for the {\sc tb} model, accounting for the uncertainties on the best fit models. Assuming shorter $\gamma$--ray events and thus, integrating over shorter time intervals $\Delta\,t$, we found no $>5\sigma$ excess, with the corresponding upper limits scaling approximately as $\sqrt{\Delta\,t}$. In particular, taking $\Delta\,t=1$~s, i.e. the time bin of the FRB, limits scale down by a factor of $\sim19$: $F_{\gamma}(40-700) < 2.5\times10^{-7}$\,erg\,cm$^{-2}$ and $< 3.7\times10^{-7}$\,erg\,cm$^{-2}$ for the {\sc pow} and {\sc tb} models, respectively.
Figure~\ref{fig:cumfluence} shows the cumulative count fluence and corresponding values in physical units for a power--law spectrum (using $\Gamma=1.84$, which gives the most conservative limit on fluence) as a function of the integration time, along with the $5\sigma$ upper limit. The most significant excess in 40--700~keV is $3.3\sigma$ around $t=14$~s. However, inspecting the individual detector profiles from Figure~\ref{fig:all32s} suggests that it is mostly due to unit 3 and due to the background modeling inaccuracy rather than a genuine signal.

Consequently, a $\gamma$--ray counterpart equal to that of FRB\,131104 cannot be ruled out for the {\sc tb} model, whereas the {\sc pow} model is rejected. Moreover, assuming a similar radio--$\gamma$ spectral slope as that found by D16, such a $\gamma$--ray counterpart is rejected with high confidence in the case of FRB\,010724, given that the Lorimer burst has a radio fluence $340\pm170$ times as high as that of FRB\,131104.

This $\gamma$--ray upper limit is the most constraining value for the radio-to-$\gamma$ fluence ratio yet obtained for an FRB: $\log_{10}{\eta}>8.2$ for {\sc pow} and $\log_{10}{\eta}>8.0$ for the {\sc tb} model. These values are $\sim250$ times higher than the value measured for FRB\,131104, $\log_{10}{\eta}=5.8\pm0.2$ (D16).
Taking the limit derived for the 1-s bin centered on the FRB, the same limits on $\eta$ become $\log_{10}{\eta}>9.5$--$9.3$ for the {\sc pow} and {\sc tb} models, respectively.
For reference, we also considered the case of a power--law with $\Gamma=2$ (Table~\ref{tab:expected}): the most conservative upper limit changes only marginally.

\begin{figure}
\begin{center}
\scalebox{1.}
{\includegraphics[width=0.5\textwidth]{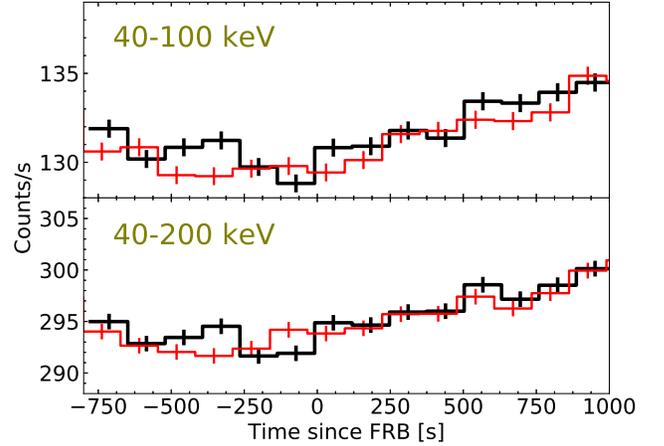}}
\caption{Mean count rates in two different energy bands with an integration time of 128~s, obtained from the 240--channel spectra continuously acquired by the GRBM units every 128~s. The FRB orbit rates (black) are shown together with the background orbit ones (red).}
\label{fig:spec128s}
\end{center}
\end{figure}

As a further check, we exploited the 240--channel spectra continuously acquired by the GRBM units in the 40--700~keV energy band every 128~s \citep{Frontera97,Guidorzi11c} to investigate the possible presence of a signal in two energy bands: $40$--$100$ and $40$--$200$~keV, respectively. Figure~\ref{fig:spec128s} displays the corresponding 128-s light curves averaged over the four units (black) along with the analogous data of the background orbit (red). Although the poorer temporal resolution does not allow us to model the background as accurately as we did for the 1-s ratemeters, yet the comparison between the FRB orbit and the background orbit rates excludes the possibility of a relatively soft signal associated with the FRB.
%

\section{Discussion}
\label{Sec:Disc}
We rule out a $\gamma$--ray transient event that is similar to Swift~J0644.5-5111 for FRB\,131104 (cf. D16). This is however assuming a power--law spectrum; the softer case of a thermal bremsstrahlung is only marginally excluded. A $\gamma$--ray signal, rescaled to the much higher radio fluence of the Lorimer burst, is also ruled out, regardless of the spectral mode. Therefore, if the association found by D16 is true, the Lorimer burst is intrinsically different and much more $\gamma$--ray quiet. Alternatively, the Lorimer burst was associated with an analogous $\gamma$--ray transient that was more collimated than the radio emission and that was not pointing towards Earth (e.g., \citealt{Romero16}).

For FRB\,010724 the DM excess, i.e. the measured DM removed of the Galactic contribution, is ${\rm DM}_{\rm E} = 330$\,pc\,cm$^{-3}$ \citep{Petroff16} and can be expressed as the sum of different terms: the intergalactic medium (IGM), the host galaxy (HG), and the local environment surrounding the FRB source:
\begin{equation}
    {\rm DM}_{\rm E}\ =\ {\rm DM}_{\rm IGM} + \frac{{\rm DM}_{\rm HG} + {\rm DM}_{\rm loc}}{1 + z}\;.
\end{equation}
Using the relation by \citet{Zhang18a} $z\sim D_{\rm IGM}/(855~{\rm pc\,cm}^{-3})$, an upper limit on redshift of $z<0.4$ is obtained.
Simulations suggest that the contribution of the host to the observed DM ranges from an average value of 45\,pc\,cm$^{-3}$ in the case of a dwarf galaxy all the way up to $\sim142$\,pc\,cm$^{-3}$ averaging over all inclination angles for a spiral \citep{XuHan15}.

\begin{figure}
\begin{center}
\scalebox{1.}
{\includegraphics[width=0.5\textwidth]{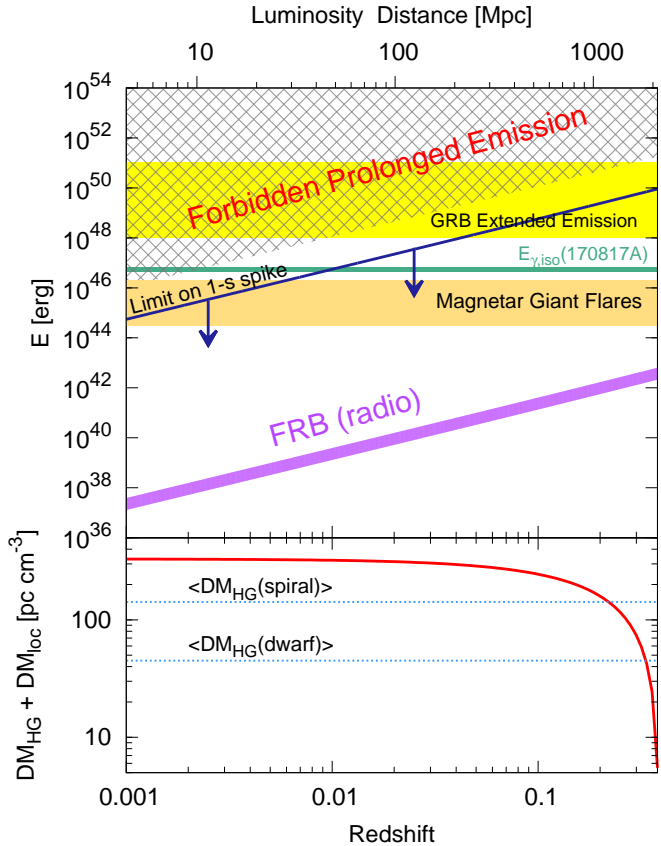}}
\caption{ {\em Top}: emitted energy of the FRB (purple) and $5\sigma$ upper limits on the 400-s prolonged emission (lower boundary of the hatched region) and on the 1-s coincident with the FRB (dark blue). For comparison, also shown are the isotropic--equivalent energies of the extended emission often associated to short GRBs (yellow), which is roughly as energetic as the short GRB itself, the sub-luminous short GRB\,170817A associated to the double NS merger GW\,170817 (green),  magnetar giant flares (orange). {\em Bottom}: DM (red) due to the combination of the host galaxy (HG) and the local environment around the source (loc), estimated subtracting the IGM contribution from the excess DM, as a function of redshift. The maximum value of $z=0.4$ corresponds to the case where all of the DM in excess of the Galactic value is due to the IGM. Also shown are the mean values of the expected DM due to the sightline within the host galaxy for two different kinds, dwarf and spiral galaxies, from simulations by \citet{XuHan15}.}
\label{fig:DMz}
\end{center}
\end{figure}

Figure~\ref{fig:DMz} summarizes our constraints for various models considered.
Assuming that the host is a dwarf, similarly to long GRBs and SLSNe as is the case for the repeater FRB\,121101 \citep{Tendulkar17,Metzger17}, and adopting for ${\rm DM}_{\rm HG}$ the typical value expected for this kind of galaxies, the constraint on redshift becomes $z\la 0.3$. Using the typical value for a spiral, i.e. about half of ${\rm DM}_{\rm E}$, the range shrinks to $z\la 0.2$. The same constraint holds assuming that half of ${\rm DM}_{\rm E}$ is due to the local environment of the FRB progenitor.
Making no assumption on ${\rm DM}_{\rm E}$, the energy of the radio pulse itself lies in the range $10^{38}< E_{\rm radio}/{\rm erg} < 6\times10^{42}$, in line with what has been argued for other FRBs, although at least 10 times more energetic than the individual bursts of the repeater FRB\,121102 \citep{Gourdji19}.

It is interesting to compare our limits on the $\gamma$--ray released energy with that of other astrophysical transients that have possible connections with FRBs.
As shown in Figure~\ref{fig:DMz}, based on our analysis, we rule out the following interpretations:
\begin{itemize}
    \item a $\gamma$--ray spike due to an extragalactic magnetar giant flare with the same radio/$\gamma$--ray fluence ratio as that of the Galactic magnetar SGR\,1806-20 (see also \citealt{Tendulkar16});
    \item short GRBs with isotropic--equivalent energies $E_{\gamma,{\rm iso}}>10^{50}$~erg, that are the majority of the observed population ($10^{49}$--$10^{52}$~erg; \citealt{Fong15});
    \item the typical $\gamma$--ray emission of a cosmological long GRB, whose isotropic-equivalent energy ranges from $\sim 10^{51}$ up to $\sim 10^{55}$~erg \citep{KWGRBcat17}.
\end{itemize}
On the other hand, we cannot reject the following scenarios:
\begin{itemize}
    \item an extragalactic giant flare as energetic as the few ones observed in our Galaxy ($3\times10^{44}$-- $2\times10^{46}$~erg; \citealt{Mazets79,Feroci01,Palmer05,Hurley05}; orange region in Figure~\ref{fig:DMz}), regardless of the radio/$\gamma$--ray spectral slope;
    \item a relatively weak short GRB ($E_{\gamma,{\rm iso}}$ in the range $10^{49}$--$10^{50}$~erg) at $z>0.04$ ($D_{\rm L}>180$~Mpc);
    \item a sub-energetic short GRB at $z>0.01$ ($D_{\rm L}>45$~Mpc), such as GRB\,170817A associated to GW\,170817, the first binary neutron star (BNS) merger detected with gravitational interferometers \citep{LIGO-Fermi17}, which had $E_{\gamma,{\rm iso}}=(5.3\pm1.0)\times10^{46}$~erg. This holds true regardless of whether the GRB is truly sub-energetic or is a typical short GRB viewed off-axis, as was the case of GRB\,170817A \citep{Margutti17gw,Margutti18a,Alexander17gw,Alexander18,Troja17nat,Granot17};
    \item the presence of the so-called extended emission (EE) that characterizes some short GRBs. This is a faint, long-lasting ($\sim$ few hundred seconds) hard X/$\gamma$--ray tail following the initial spike \citep{Lazzati01,Montanari05,Norris06}, whose isotropic-equivalent energy is in the range $10^{48}$--$10^{51}$~erg (\citealt{Kisaka17}; yellow area in Fig.~\ref{fig:DMz});
    \item the so-called low-luminosity ({\em ll}) long GRBs \citep{Kulkarni98,Campana06,Waxman07} having $E_{\gamma,{\rm iso}}\sim10^{48}$--$10^{50}$~erg. {\em ll}-GRBs probably represent a separate family \citep{Liang07,Amati07,Virgili09,WandermanPiran10,HowellCoward13} and are possibly the result of a relativistic shock breakout \citep{Nakar12}.
\end{itemize}

Further considerations on the possible association between FRBs and GRBs and the other classes mentioned above are based on the relative volumetric rates. Although the FRB rate does depend on the redshift distribution, which is essentially unknown, yet using ${\rm DM}_{\rm E}$ as a proxy for distance and assuming that the observed population is uniformily distributed within $z\la 1$, a rate of $\la 10^4$~Gpc$^{-3}$\,yr$^{-1}$ is obtained \citep{Rane16,Scholz16,Crawford16},
which is not too different from the local one of Ib/c SNe \citep{Dahlen04,Li11,Cappellaro15}.
Our conjecture that long (i.e., not {\em ll}-) cosmological GRBs are not associated with the Lorimer burst is consistent with the different local volumetric rate of GRBs: $1.3^{+0.6}_{-0.7}\,f_b^{-1}$\,Gpc$^{-3}$\,yr$^{-1}$ \citep{WandermanPiran10} is $\sim10$--$100$ times lower than that of FRBs even after accounting for a $f_b^{-1}=100$ beaming-corrected fraction. Somewhat higher values are obtained for short GRBs, whose rate is $4$--$8\,f_b^{-1}$\,Gpc$^{-3}$\,yr$^{-1}$ \citep{Coward12,WandermanPiran15}.
Although poorly constrained, the BNS rate of $1540_{-1220}^{+3200}$\,Gpc$^{-3}$\,yr$^{-1}$ from the recent gravitational wave observations \citep{LIGO-Virgo17_prl} is not too far from that of FRBs.
While a {\em ll}-GRB associated with the Lorimer burst is not excluded, their estimated local rate ($\sim200$--$300\,f_b^{-1}$\,Gpc$^{-3}$\,yr$^{-1}$ with a lower $f_b^{-1}$ than for long GRBs; \citealt{Soderberg06d,Liang07}) is also significantly lower than that of FRBs. 



\section{Conclusions}
\label{Sec:Conc}
Motivated by the possible and controversial detection of a transient hard X/$\gamma$--ray source positionally and temporally compatible with FRB\,131104, we considered the brightest Lorimer FRB and examined the data of {\em BeppoSAX}/GRBM, one of the most sensitive instruments that in principle could have detected any simultaneous emission in the keV--MeV band. A previous upper limit on the $\gamma$--ray fluence of $2\times10^{-7}$~erg~cm$^{-2}$ in the Konus/WIND passband was already adopted in the past \citep{Tendulkar16}. However, this value is not the outcome of any specific data analysis, but merely corresponds to the lowest fluence detected in the Konus/WIND GRB catalog: this is because (i) the detection efficiency in the lowest end of the fluence distribution is likely to be significantly below 1; (ii) the typical duration of a GRB is one order of magnitude less than the transient signal possibly associated with FRB\,131104; (iii) no specific analysis in that energy band at the time of the Lorimer burst has been reported yet. Based on our analysis we rule out a $\gamma$--ray counterpart for the Lorimer burst with the same radio/$\gamma$--ray fluence ratio as that of FRB\,131104. Furthermore, the combination of a very high specific flux and a relatively low DM strongly suggests this FRB to be a relatively nearby FRB. We used our fluence upper limits to investigate a possible relation of FRB sources with other classes of transient phenomena.
We rule out an extragalactic magnetar flare analogous to that of the Galactic source SGR\,1806-20, as well as a typical long GRB. Instead, a low-luminosity GRB or a sub-energetic short GRB (such as a standard short GRB viewed off-axis, as was probably the case of GRB\,170817A associated to the first binary NS merger) cannot be excluded. These results, along with the comparison of the relative volumetric rates of the corresponding populations, do not rule out the possible association of FRBs with either binary NS mergers or with {\em ll}-GRBs, which, however, looks incompatible with a naive one-to-one correspondence.

\bigskip
We thank the anonymous referee for the detailed comments that improved the paper. We thank Marcello Giroletti for his careful reading and comments. Support for this work was provided by Universit\`a di Ferrara through grant FIR~2018 ``A Broad-band study of Cosmic Gamma-Ray Burst Prompt and Afterglow Emission" (PI Guidorzi). This work is part of the INAF PRIN SKA-CTA 2016.
L.A. acknowledges financial contribution from the agreement ASI-INAF n.2017-14-H.O.


\begin{thebibliography}{}
\expandafter\ifx\csname natexlab\endcsname\relax\def\natexlab#1{#1}\fi

\bibitem[{{Aartsen} {et~al.}(2018){Aartsen}, {Ackermann}, {Adams}, {Aguilar},
  {Ahlers}, {Ahrens}, {Samarai}, {Altmann}, {Andeen}, {Anderson}, \&
  et~al.}]{IceCube18_FRB}
{Aartsen}, M.~G., {Ackermann}, M., {Adams}, J., {et~al.} 2018, \apj, 857, 117

\bibitem[{{Abbott} {et~al.}(2017{\natexlab{a}}){Abbott}, {Abbott}, {Abbott},
  {Acernese}, {Ackley}, {Adams}, {Adams}, {Addesso}, {Adhikari}, {Adya}, \&
  et~al.}]{LIGO-Fermi17}
{Abbott}, B.~P., {Abbott}, R., {Abbott}, T.~D., {et~al.} 2017{\natexlab{a}},
  \apjl, 848, L13

\bibitem[{{Abbott} {et~al.}(2017{\natexlab{b}}){Abbott}, {Abbott}, {Abbott},
  {Acernese}, {Ackley}, {Adams}, {Adams}, {Addesso}, {Adhikari}, {Adya}, \&
  et~al.}]{LIGO-Virgo17_prl}
---. 2017{\natexlab{b}}, Physical Review Letters, 119, 161101

\bibitem[{{Albert} {et~al.}(2019){Albert}, {Andr{\'e}}, {Anghinolfi}, {Anton},
  {Ardid}, {Aubert}, {Aublin}, {Avgitas}, {Baret}, {Barrios-Mart{\'{\i}}},
  {Basa}, {Belhorma}, {Bertin}, {Biagi}, {Bormuth}, {Boumaaza}, {Bourret},
  {Bouwhuis}, {Br{\^a}nza{\c s}}, {Bruijn}, {Brunner}, {Busto}, {Capone},
  {Caramete}, {Carr}, {Celli}, {Chabab}, {Cherkaoui El Moursli}, {Chiarusi},
  {Circella}, {Coelho}, {Coleiro}, {Colomer}, {Coniglione}, {Costantini},
  {Coyle}, {Creusot}, {D{\'{\i}}az}, {Deschamps}, {Distefano}, {Di Palma},
  {Domi}, {Donzaud}, {Dornic}, {Drouhin}, {Eberl}, {El Bojaddaini}, {El
  Khayati}, {Els{\"a}sser}, {Enzenh{\"o}fer}, {Ettahiri}, {Fassi}, {Felis},
  {Fermani}, {Ferrara}, {Fusco}, {Gay}, {Glotin}, {Gr{\'e}goire},
  {Gracia-Ruiz}, {Graf}, {Hallmann}, {van Haren}, {Heijboer}, {Hello},
  {Hern{\'a}ndez-Rey}, {H{\"o}{\ss}l}, {Hofest{\"a}dt}, {Illuminati}, {James},
  {de Jong}, {Jongen}, {Kadler}, {Kalekin}, {Katz}, {Kouchner}, {Kreter},
  {Kreykenbohm}, {Kulikovskiy}, {Lachaud}, {Lahmann}, {Lef{\`e}vre}, {Leonora},
  {Levi}, {Lotze}, {Loucatos}, {Marcelin}, {Margiotta}, {Marinelli},
  {Mart{\'{\i}}nez-Mora}, {Mele}, {Melis}, {Migliozzi}, {Moussa}, {Navas},
  {Nezri}, {Nu{\~n}ez}, {Organokov}, {P{\v a}v{\v a}la{\c s}}, {Pellegrino},
  {Piattelli}, {Popa}, {Pradier}, {Quinn}, {Racca}, {Randazzo}, {Riccobene},
  {S{\'a}nchez-Losa}, {Salda{\~n}a}, {Salvadori}, {Samtleben}, {Sanguineti},
  {Sapienza}, {Sch{\"u}ssler}, {Spurio}, {Stolarczyk}, {Taiuti}, {Tayalati},
  {Trovato}, {Turpin}, {Vallage}, {Van Elewyck}, {Versari}, {Vivolo}, {Wilms},
  {Zaborov}, {Zornoza}, \& {Z{\'u}{\~n}iga}}]{ANTARES19}
{Albert}, A., {Andr{\'e}}, M., {Anghinolfi}, M., {et~al.} 2019, \mnras, 482,
  184

\bibitem[{{Alexander} {et~al.}(2017){Alexander}, {Berger}, {Fong}, {Williams},
  {Guidorzi}, {Margutti}, {Metzger}, {Annis}, {Blanchard}, {Brout}, {Brown},
  {Chen}, {Chornock}, {Cowperthwaite}, {Drout}, {Eftekhari}, {Frieman}, {Holz},
  {Nicholl}, {Rest}, {Sako}, {Soares-Santos}, \& {Villar}}]{Alexander17gw}
{Alexander}, K.~D., {Berger}, E., {Fong}, W., {et~al.} 2017, \apjl, 848, L21

\bibitem[{{Alexander} {et~al.}(2018){Alexander}, {Margutti}, {Blanchard},
  {Fong}, {Berger}, {Hajela}, {Eftekhari}, {Chornock}, {Cowperthwaite},
  {Giannios}, {Guidorzi}, {Kathirgamaraju}, {MacFadyen}, {Metzger}, {Nicholl},
  {Sironi}, {Villar}, {Williams}, {Xie}, \& {Zrake}}]{Alexander18}
{Alexander}, K.~D., {Margutti}, R., {Blanchard}, P.~K., {et~al.} 2018, \apjl,
  863, L18

\bibitem[{{Amati} {et~al.}(2007){Amati}, {Della Valle}, {Frontera}, {Malesani},
  {Guidorzi}, {Montanari}, \& {Pian}}]{Amati07}
{Amati}, L., {Della Valle}, M., {Frontera}, F., {et~al.} 2007, A\&A, 463, 913

\bibitem[{{Atwood} {et~al.}(2009){Atwood}, {Abdo}, {Ackermann}, {Althouse},
  {Anderson}, {Axelsson}, {Baldini}, {Ballet}, {Band}, {Barbiellini},
  {Bartelt}, {Bastieri}, {Baughman}, {Bechtol}, {B{\'e}d{\'e}r{\`e}de},
  {Bellardi}, {Bellazzini}, {Berenji}, {Bignami}, {Bisello}, {Bissaldi},
  {Blandford}, {Bloom}, {Bogart}, {Bonamente}, {Bonnell}, {Borgland },
  {Bouvier}, {Bregeon}, {Brez}, {Brigida}, {Bruel}, {Burnett}, {Busetto},
  {Caliandro}, {Cameron}, {Caraveo}, {Carius}, {Carlson}, {Casandjian},
  {Cavazzuti}, {Ceccanti}, {Cecchi}, {Charles}, {Chekhtman}, {Cheung},
  {Chiang}, {Chipaux}, {Cillis}, {Ciprini}, {Claus}, {Cohen-Tanugi},
  {Condamoor}, {Conrad}, {Corbet}, {Corucci}, {Costamante}, {Cutini}, {Davis},
  {Decotigny}, {DeKlotz}, {Dermer}, {de Angelis}, {Digel}, {do Couto e Silva},
  {Drell}, {Dubois}, {Dumora}, {Edmonds}, {Fabiani}, {Farnier}, {Favuzzi},
  {Flath}, {Fleury}, {Focke}, {Funk}, {Fusco}, {Gargano}, {Gasparrini},
  {Gehrels}, {Gentit}, {Germani}, {Giebels}, {Giglietto}, {Giommi}, {Giordano},
  {Glanzman}, {Godfrey}, {Grenier}, {Grondin}, {Grove}, {Guillemot}, {Guiriec},
  {Haller}, {Harding}, {Hart}, {Hays}, {Healey}, {Hirayama}, {Hjalmarsdotter},
  {Horn}, {Hughes}, {J{\'o}hannesson}, {Johansson}, {Johnson}, {Johnson},
  {Johnson}, {Johnson}, {Kamae}, {Katagiri}, {Kataoka}, {Kavelaars}, {Kawai},
  {Kelly}, {Kerr}, {Klamra}, {Kn{\"o}dlseder}, {Kocian}, {Komin}, {Kuehn},
  {Kuss}, {Landriu}, {Latronico}, {Lee}, {Lee}, {Lemoine-Goumard}, {Lionetto},
  {Longo}, {Loparco}, {Lott}, {Lovellette}, {Lubrano}, {Madejski}, {Makeev},
  {Marangelli}, {Massai}, {Mazziotta}, {McEnery}, {Menon}, {Meurer},
  {Michelson}, {Minuti}, {Mirizzi}, {Mitthumsiri}, {Mizuno}, {Moiseev},
  {Monte}, {Monzani}, {Moretti}, {Morselli}, {Moskalenko}, {Murgia},
  {Nakamori}, {Nishino}, {Nolan}, {Norris}, {Nuss}, {Ohno}, {Ohsugi}, {Omodei},
  {Orlando}, {Ormes}, {Paccagnella}, {Paneque}, {Panetta}, {Parent}, {Pearce},
  {Pepe}, {Perazzo}, {Pesce-Rollins}, {Picozza}, {Pieri}, {Pinchera}, {Piron},
  {Porter}, {Poupard}, {Rain{\`o}}, {Rando}, {Rapposelli}, {Razzano}, {Reimer},
  {Reimer}, {Reposeur}, {Reyes}, {Ritz}, {Rochester}, {Rodriguez}, {Romani},
  {Roth}, {Russell}, {Ryde}, {Sabatini}, {Sadrozinski}, {Sanchez}, {Sand er},
  {Sapozhnikov}, {Parkinson}, {Scargle}, {Schalk}, {Scolieri}, {Sgr{\`o}},
  {Share}, {Shaw}, {Shimokawabe}, {Shrader}, {Sierpowska-Bartosik}, {Siskind},
  {Smith}, {Smith}, {Spandre}, {Spinelli}, {Starck}, {Stephens}, {Strickman},
  {Strong}, {Suson}, {Tajima}, {Takahashi}, {Takahashi}, {Tanaka}, {Tenze},
  {Tether}, {Thayer}, {Thayer}, {Thompson}, {Tibaldo}, {Tibolla}, {Torres},
  {Tosti}, {Tramacere}, {Turri}, {Usher}, {Vilchez}, {Vitale}, {Wang},
  {Watters}, {Winer}, {Wood}, {Ylinen}, \& {Ziegler}}]{Atwood09}
{Atwood}, W.~B., {Abdo}, A.~A., {Ackermann}, M., {et~al.} 2009, \apj, 697, 1071

\bibitem[{{Bannister} {et~al.}(2019){Bannister}, {Deller}, {Phillips},
  {Macquart}, {Prochaska}, {Tejos}, {Ryder}, {Sadler}, {Shannon}, \&
  {Simha}}]{Bannister19}
{Bannister}, K.~W., {Deller}, A.~T., {Phillips}, C., {et~al.} 2019, arXiv
  e-prints, arXiv:1906.11476

\bibitem[{{Barthelmy} {et~al.}(2005){Barthelmy}, {Barbier}, {Cummings},
  {Fenimore}, {Gehrels}, {Hullinger}, {Krimm}, {Markwardt}, {Palmer},
  {Parsons}, {Sato}, {Suzuki}, {Takahashi}, {Tashiro}, \&
  {Tueller}}]{Barthelmy05}
{Barthelmy}, S.~D., {Barbier}, L.~M., {Cummings}, J.~R., {et~al.} 2005, Space
  Sci. Rev., 120, 143

\bibitem[{{Boella} {et~al.}(1997){Boella}, {Butler}, {Perola}, {Piro},
  {Scarsi}, \& {Bleeker}}]{Boella97}
{Boella}, G., {Butler}, R.~C., {Perola}, G.~C., {et~al.} 1997, \aaps, 122, 299

\bibitem[{{Bucciantini} {et~al.}(2007){Bucciantini}, {Quataert}, {Arons},
  {Metzger}, \& {Thompson}}]{Bucciantini07}
{Bucciantini}, N., {Quataert}, E., {Arons}, J., {Metzger}, B.~D., \&
  {Thompson}, T.~A. 2007, \mnras, 380, 1541

\bibitem[{{Caleb} {et~al.}(2019){Caleb}, {Stappers}, {Rajwade}, \&
  {Flynn}}]{Caleb19}
{Caleb}, M., {Stappers}, B.~W., {Rajwade}, K., \& {Flynn}, C. 2019, \mnras,
  484, 5500

\bibitem[{{Campana} {et~al.}(2006){Campana}, {Mangano}, {Blustin}, {Brown},
  {Burrows}, {Chincarini}, {Cummings}, {Cusumano}, {Della Valle}, {Malesani},
  {M{\'e}sz{\'a}ros}, {Nousek}, {Page}, {Sakamoto}, {Waxman}, {Zhang}, {Dai},
  {Gehrels}, {Immler}, {Marshall}, {Mason}, {Moretti}, {O'Brien}, {Osborne},
  {Page}, {Romano}, {Roming}, {Tagliaferri}, {Cominsky}, {Giommi}, {Godet},
  {Kennea}, {Krimm}, {Angelini}, {Barthelmy}, {Boyd}, {Palmer}, {Wells}, \&
  {White}}]{Campana06}
{Campana}, S., {Mangano}, V., {Blustin}, A.~J., {et~al.} 2006, \nat, 442, 1008

\bibitem[{{Cappellaro} {et~al.}(2015){Cappellaro}, {Botticella}, {Pignata},
  {Grado}, {Greggio}, {Limatola}, {Vaccari}, {Baruffolo}, {Benetti}, {Bufano},
  {Capaccioli}, {Cascone}, {Covone}, {De Cicco}, {Falocco}, {Della Valle},
  {Jarvis}, {Marchetti}, {Napolitano}, {Paolillo}, {Pastorello}, {Radovich},
  {Schipani}, {Spiro}, {Tomasella}, \& {Turatto}}]{Cappellaro15}
{Cappellaro}, E., {Botticella}, M.~T., {Pignata}, G., {et~al.} 2015, \aap, 584,
  A62

\bibitem[{{Chatterjee} {et~al.}(2017){Chatterjee}, {Law}, {Wharton},
  {Burke-Spolaor}, {Hessels}, {Bower}, {Cordes}, {Tendulkar}, {Bassa},
  {Demorest}, {Butler}, {Seymour}, {Scholz}, {Abruzzo}, {Bogdanov}, {Kaspi},
  {Keimpema}, {Lazio}, {Marcote}, {McLaughlin}, {Paragi}, {Ransom}, {Rupen},
  {Spitler}, \& {van Langevelde}}]{Chatterjee17}
{Chatterjee}, S., {Law}, C.~J., {Wharton}, R.~S., {et~al.} 2017, \nat, 541, 58

\bibitem[{{CHIME/FRB Collaboration} {et~al.}(2019){CHIME/FRB Collaboration},
  {Amiri}, {Bandura}, {Bhardwaj}, {Boubel}, {Boyce}, {Boyle}, {.~Brar},
  {Burhanpurkar}, {Cassanelli}, {Chawla}, {Cliche}, {Cubranic}, {Deng},
  {Denman}, {Dobbs}, {Fandino}, {Fonseca}, {Gaensler}, {Gilbert}, {Gill},
  {Giri}, {Good}, {Halpern}, {Hanna}, {Hill}, {Hinshaw}, {H{\"o}fer},
  {Josephy}, {Kaspi}, {Landecker}, {Lang}, {Lin}, {Masui}, {Mckinven},
  {Mena-Parra}, {Merryfield}, {Michilli}, {Milutinovic}, {Moatti}, {Naidu},
  {Newburgh}, {Ng}, {Patel}, {Pen}, {Pinsonneault-Marotte}, {Pleunis},
  {Rafiei-Ravandi}, {Rahman}, {Ransom}, {Renard}, {Scholz}, {Shaw}, {Siegel},
  {Smith}, {Stairs}, {Tendulkar}, {Tretyakov}, {Vanderlinde}, \&
  {Yadav}}]{CHIME19a}
{CHIME/FRB Collaboration}, {Amiri}, M., {Bandura}, K., {et~al.} 2019, \nat,
  566, 235

\bibitem[{{Cordes} \& {Wasserman}(2016)}]{CordesWassermann16}
{Cordes}, J.~M., \& {Wasserman}, I. 2016, \mnras, 457, 232

\bibitem[{{Coward} {et~al.}(2012){Coward}, {Howell}, {Piran}, {Stratta},
  {Branchesi}, {Bromberg}, {Gendre}, {Burman}, \& {Guetta}}]{Coward12}
{Coward}, D.~M., {Howell}, E.~J., {Piran}, T., {et~al.} 2012, \mnras, 425, 2668

\bibitem[{{Crawford} {et~al.}(2016){Crawford}, {Rane}, {Tran}, {Rolph},
  {Lorimer}, \& {Ridley}}]{Crawford16}
{Crawford}, F., {Rane}, A., {Tran}, L., {et~al.} 2016, \mnras, 460, 3370

\bibitem[{{Dahlen} {et~al.}(2004){Dahlen}, {Strolger}, {Riess}, {Mobasher},
  {Chary}, {Conselice}, {Ferguson}, {Fruchter}, {Giavalisco}, {Livio}, {Madau},
  {Panagia}, \& {Tonry}}]{Dahlen04}
{Dahlen}, T., {Strolger}, L.-G., {Riess}, A.~G., {et~al.} 2004, \apj, 613, 189

\bibitem[{{DeLaunay} {et~al.}(2016){DeLaunay}, {Fox}, {Murase},
  {M{\'e}sz{\'a}ros}, {Keivani}, {Messick}, {Mostaf{\'a}}, {Oikonomou}, {Te{\v
  s}i{\'c}}, \& {Turley}}]{DeLaunay16}
{DeLaunay}, J.~J., {Fox}, D.~B., {Murase}, K., {et~al.} 2016, \apjl, 832, L1

\bibitem[{{Eftekhari} \& {Berger}(2017)}]{EftekhariBerger17}
{Eftekhari}, T., \& {Berger}, E. 2017, \apj, 849, 162

\bibitem[{{Fan} \& {Xu}(2006)}]{FanXu06}
{Fan}, Y.-Z., \& {Xu}, D. 2006, \mnras, 372, L19

\bibitem[{{Feroci} {et~al.}(2001){Feroci}, {Hurley}, {Duncan}, \&
  {Thompson}}]{Feroci01}
{Feroci}, M., {Hurley}, K., {Duncan}, R.~C., \& {Thompson}, C. 2001, \apj, 549,
  1021

\bibitem[{{Fong} {et~al.}(2015){Fong}, {Berger}, {Margutti}, \&
  {Zauderer}}]{Fong15}
{Fong}, W., {Berger}, E., {Margutti}, R., \& {Zauderer}, B.~A. 2015, \apj, 815,
  102

\bibitem[{{Frontera} {et~al.}(1997){Frontera}, {Costa}, {dal Fiume}, {Feroci},
  {Nicastro}, {Orlandini}, {Palazzi}, \& {Zavattini}}]{Frontera97}
{Frontera}, F., {Costa}, E., {dal Fiume}, D., {et~al.} 1997, \aaps, 122, 357

\bibitem[{{Frontera} {et~al.}(2009){Frontera}, {Guidorzi}, {Montanari},
  {Rossi}, {Costa}, {Feroci}, {Calura}, {Rapisarda}, {Amati}, {Carturan},
  {Cinti}, {Fiume}, {Nicastro}, \& {Orlandini}}]{Frontera09}
{Frontera}, F., {Guidorzi}, C., {Montanari}, E., {et~al.} 2009, ApJS, 180, 192

\bibitem[{{Gao} \& {Zhang}(2017)}]{GaoZhang17}
{Gao}, H., \& {Zhang}, B. 2017, \apjl, 835, L21

\bibitem[{{Gehrels} {et~al.}(2004){Gehrels}, {Chincarini}, {Giommi}, {Mason},
  {Nousek}, {Wells}, {White}, {Barthelmy}, {Burrows}, {Cominsky}, {Hurley},
  {Marshall}, {M{\'e}sz{\'a}ros}, {Roming}, {Angelini}, {Barbier}, {Belloni},
  {Campana}, {Caraveo}, {Chester}, {Citterio}, {Cline}, {Cropper}, {Cummings},
  {Dean}, {Feigelson}, {Fenimore}, {Frail}, {Fruchter}, {Garmire}, {Gendreau},
  {Ghisellini}, {Greiner}, {Hill}, {Hunsberger}, {Krimm}, {Kulkarni}, {Kumar},
  {Lebrun}, {Lloyd-Ronning}, {Markwardt}, {Mattson}, {Mushotzky}, {Norris},
  {Osborne}, {Paczynski}, {Palmer}, {Park}, {Parsons}, {Paul}, {Rees},
  {Reynolds}, {Rhoads}, {Sasseen}, {Schaefer}, {Short}, {Smale}, {Smith},
  {Stella}, {Tagliaferri}, {Takahashi}, {Tashiro}, {Townsley}, {Tueller},
  {Turner}, {Vietri}, {Voges}, {Ward}, {Willingale}, {Zerbi}, \&
  {Zhang}}]{Gehrels04}
{Gehrels}, N., {Chincarini}, G., {Giommi}, P., {et~al.} 2004, ApJ, 611, 1005

\bibitem[{{Gompertz} {et~al.}(2013){Gompertz}, {O'Brien}, {Wynn}, \&
  {Rowlinson}}]{Gompertz13}
{Gompertz}, B.~P., {O'Brien}, P.~T., {Wynn}, G.~A., \& {Rowlinson}, A. 2013,
  \mnras, 431, 1745

\bibitem[{{Gourdji} {et~al.}(2019){Gourdji}, {Michilli}, {Spitler}, {Hessels},
  {Seymour}, {Cordes}, \& {Chatterjee}}]{Gourdji19}
{Gourdji}, K., {Michilli}, D., {Spitler}, L.~G., {et~al.} 2019, \apj, 877, L19

\bibitem[{{Granot} {et~al.}(2017){Granot}, {Guetta}, \& {Gill}}]{Granot17}
{Granot}, J., {Guetta}, D., \& {Gill}, R. 2017, \apjl, 850, L24

\bibitem[{{Guidorzi} {et~al.}(2011){Guidorzi}, {Lacapra}, {Frontera},
  {Montanari}, {Amati}, {Calura}, {Nicastro}, \& {Orlandini}}]{Guidorzi11c}
{Guidorzi}, C., {Lacapra}, M., {Frontera}, F., {et~al.} 2011, A\&A, 526, A49

\bibitem[{{H.E.S.S.~Collaboration} {et~al.}(2017){H.E.S.S.~Collaboration},
  {Abdalla}, {Abramowski}, {Aharonian}, {Ait Benkhali}, {Akhperjanian},
  {Andersson}, {Ang{\"u}ner}, {Arakawa}, {Arrieta}, \& et~al.}]{HESS17}
{H.E.S.S.~Collaboration}, {Abdalla}, H., {Abramowski}, A., {et~al.} 2017, \aap,
  597, A115

\bibitem[{{Howell} \& {Coward}(2013)}]{HowellCoward13}
{Howell}, E.~J., \& {Coward}, D.~M. 2013, \mnras, 428, 167

\bibitem[{{Hurley} {et~al.}(2005){Hurley}, {Boggs}, {Smith}, {Duncan}, {Lin},
  {Zoglauer}, {Krucker}, {Hurford}, {Hudson}, {Wigger}, {Hajdas}, {Thompson},
  {Mitrofanov}, {Sanin}, {Boynton}, {Fellows}, {von Kienlin}, {Lichti}, {Rau},
  \& {Cline}}]{Hurley05}
{Hurley}, K., {Boggs}, S.~E., {Smith}, D.~M., {et~al.} 2005, \nat, 434, 1098

\bibitem[{{Jager} {et~al.}(1997){Jager}, {Mels}, {Brinkman}, {Galama},
  {Goulooze}, {Heise}, {Lowes}, {Muller}, {Naber}, {Rook}, {Schuurhof},
  {Schuurmans}, \& {Wiersma}}]{Jager97}
{Jager}, R., {Mels}, W.~A., {Brinkman}, A.~C., {et~al.} 1997, \aaps, 125, 557

\bibitem[{{Katz}(2014)}]{Katz14}
{Katz}, J.~I. 2014, \prd, 89, 103009

\bibitem[{{Katz}(2016)}]{Katz16b}
---. 2016, \apj, 818, 19

\bibitem[{{Katz}(2018)}]{Katz18rev}
---. 2018, Progress in Particle and Nuclear Physics, 103, 1

\bibitem[{{Kisaka} {et~al.}(2017){Kisaka}, {Ioka}, \& {Sakamoto}}]{Kisaka17}
{Kisaka}, S., {Ioka}, K., \& {Sakamoto}, T. 2017, \apj, 846, 142

\bibitem[{{Kocevski} {et~al.}(2003){Kocevski}, {Ryde}, \& {Liang}}]{Kocevski03}
{Kocevski}, D., {Ryde}, F., \& {Liang}, E. 2003, \apj, 596, 389

\bibitem[{{Kulkarni} {et~al.}(1998){Kulkarni}, {Frail}, {Wieringa}, {Ekers},
  {Sadler}, {Wark}, {Higdon}, {Phinney}, \& {Bloom}}]{Kulkarni98}
{Kulkarni}, S.~R., {Frail}, D.~A., {Wieringa}, M.~H., {et~al.} 1998, \nat, 395,
  663

\bibitem[{{Kumar} \& {Zhang}(2015)}]{KumarZhang15rev}
{Kumar}, P., \& {Zhang}, B. 2015, Phys. Rep., 561, 1

\bibitem[{{Lazzati} {et~al.}(2001){Lazzati}, {Ramirez-Ruiz}, \&
  {Ghisellini}}]{Lazzati01}
{Lazzati}, D., {Ramirez-Ruiz}, E., \& {Ghisellini}, G. 2001, \aap, 379, L39

\bibitem[{{Li} {et~al.}(2011){Li}, {Chornock}, {Leaman}, {Filippenko},
  {Poznanski}, {Wang}, {Ganeshalingam}, \& {Mannucci}}]{Li11}
{Li}, W., {Chornock}, R., {Leaman}, J., {et~al.} 2011, \mnras, 412, 1473

\bibitem[{{Liang} {et~al.}(2007){Liang}, {Zhang}, {Virgili}, \&
  {Dai}}]{Liang07}
{Liang}, E., {Zhang}, B., {Virgili}, F., \& {Dai}, Z.~G. 2007, \apj, 662, 1111

\bibitem[{{Lorimer} {et~al.}(2007){Lorimer}, {Bailes}, {McLaughlin},
  {Narkevic}, \& {Crawford}}]{Lorimer07}
{Lorimer}, D.~R., {Bailes}, M., {McLaughlin}, M.~A., {Narkevic}, D.~J., \&
  {Crawford}, F. 2007, Science, 318, 777

\bibitem[{{Lyutikov} {et~al.}(2016){Lyutikov}, {Burzawa}, \&
  {Popov}}]{Lyutikov16a}
{Lyutikov}, M., {Burzawa}, L., \& {Popov}, S.~B. 2016, \mnras, 462, 941

\bibitem[{{Macquart} \& {Ekers}(2018)}]{MacquartEkers18a}
{Macquart}, J.-P., \& {Ekers}, R.~D. 2018, \mnras, 474, 1900

\bibitem[{{MAGIC Collaboration} {et~al.}(2018){MAGIC Collaboration}, {Acciari},
  {Ansoldi}, {Antonelli}, {Arbet Engels}, {Arcaro}, {Baack}, {Babi{\'c}},
  {Banerjee}, {Bangale}, {Barres de Almeida}, {Barrio}, {Becerra Gonz{\'a}lez},
  {Bednarek}, {Bernardini}, {Berti}, {Besenrieder}, {Bhattacharyya},
  {Bigongiari}, {Biland}, {Blanch}, {Bonnoli}, {Carosi}, {Ceribella},
  {Chatterjee}, {Colak}, {Colin}, {Colombo}, {Contreras}, {Cortina}, {Covino},
  {Cumani}, {D'Elia}, {Da Vela}, {Dazzi}, {De Angelis}, {De Lotto}, {Delfino},
  {Delgado}, {Di Pierro}, {Dom{\'\i}nguez}, {Dominis Prester}, {Dorner},
  {Doro}, {Einecke}, {Elsaesser}, {Fallah Ramazani}, {Fattorini},
  {Fern{\'a}ndez-Barral}, {Ferrara}, {Fidalgo}, {Foffano}, {Fonseca}, {Font},
  {Fruck}, {Gallozzi}, {Garc{\'\i}a L{\'o}pez}, {Garczarczyk}, {Gaug},
  {Giammaria}, {Godinovi{\'c}}, {Guberman}, {Hadasch}, {Hahn}, {Hassan},
  {Herrera}, {Hoang}, {Hrupec}, {Inoue}, {Ishio}, {Iwamura}, {Kubo}, {Kushida},
  {Kuve{\v{z}}di{\'c}}, {Lamastra}, {Lelas}, {Leone}, {Lindfors}, {Lombardi},
  {Longo}, {L{\'o}pez}, {L{\'o}pez-Oramas}, {Maggio}, {Majumdar}, {Makariev},
  {Maneva}, {Manganaro}, {Mannheim}, {Maraschi}, {Mariotti}, {Mart{\'\i}nez},
  {Masuda}, {Mazin}, {Minev}, {Miranda}, {Mirzoyan}, {Molina}, {Moralejo},
  {Moreno}, {Moretti}, {Neustroev}, {Niedzwiecki}, {Nievas Rosillo}, {Nigro},
  {Nilsson}, {Ninci}, {Nishijima}, {Noda}, {Nogu{\'e}s}, {Paiano}, {Palacio},
  {Paneque}, {Paoletti}, {Paredes}, {Pedaletti}, {Pe{\~n}il}, {Peresano},
  {Persic}, {Prada Moroni}, {Prand ini}, {Puljak}, {Garcia}, {Rhode},
  {Rib{\'o}}, {Rico}, {Righi}, {Rugliancich}, {Saha}, {Saito}, {Satalecka},
  {Schweizer}, {Sitarek}, {{\v{S}}nidari{\'c}}, {Sobczynska}, {Somero},
  {Stamerra}, {Strzys}, {Suri{\'c}}, {Tavecchio}, {Temnikov}, {Terzi{\'c}},
  {Teshima}, {Torres-Alb{\`a}}, {Tsujimoto}, {Vanzo}, {Vazquez Acosta}, {Vovk},
  {Ward}, {Will}, {Zari{\'c}}, {Marcote}, {Spitler}, {Hessels}, {Kashiyama},
  {Murase}, {Bosch-Ramon}, {Michilli}, \& {Seymour}}]{MAGIC18}
{MAGIC Collaboration}, {Acciari}, V.~A., {Ansoldi}, S., {et~al.} 2018, \mnras,
  481, 2479

\bibitem[{{Margutti} {et~al.}(2017){Margutti}, {Berger}, {Fong}, {Guidorzi},
  {Alexander}, {Metzger}, {Blanchard}, {Cowperthwaite}, {Chornock},
  {Eftekhari}, {Nicholl}, {Villar}, {Williams}, {Annis}, {Brown}, {Chen},
  {Doctor}, {Frieman}, {Holz}, {Sako}, \& {Soares-Santos}}]{Margutti17gw}
{Margutti}, R., {Berger}, E., {Fong}, W., {et~al.} 2017, \apjl, 848, L20

\bibitem[{{Margutti} {et~al.}(2018){Margutti}, {Alexander}, {Xie}, {Sironi},
  {Metzger}, {Kathirgamaraju}, {Fong}, {Blanchard}, {Berger}, {MacFadyen},
  {Giannios}, {Guidorzi}, {Hajela}, {Chornock}, {Cowperthwaite}, {Eftekhari},
  {Nicholl}, {Villar}, {Williams}, \& {Zrake}}]{Margutti18a}
{Margutti}, R., {Alexander}, K.~D., {Xie}, X., {et~al.} 2018, \apjl, 856, L18

\bibitem[{{Mazets} {et~al.}(1979){Mazets}, {Golentskii}, {Ilinskii}, {Aptekar},
  \& {Guryan}}]{Mazets79}
{Mazets}, E.~P., {Golentskii}, S.~V., {Ilinskii}, V.~N., {Aptekar}, R.~L., \&
  {Guryan}, I.~A. 1979, \nat, 282, 587

\bibitem[{{Meegan} {et~al.}(2009){Meegan}, {Lichti}, {Bhat}, {Bissaldi},
  {Briggs}, {Connaughton}, {Diehl}, {Fishman}, {Greiner}, {Hoover}, {van der
  Horst}, {von Kienlin}, {Kippen}, {Kouveliotou}, {McBreen}, {Paciesas},
  {Preece}, {Steinle}, {Wallace}, {Wilson}, \& {Wilson-Hodge}}]{Meegan09}
{Meegan}, C., {Lichti}, G., {Bhat}, P.~N., {et~al.} 2009, ApJ, 702, 791

\bibitem[{{Metzger} {et~al.}(2017){Metzger}, {Berger}, \&
  {Margalit}}]{Metzger17}
{Metzger}, B.~D., {Berger}, E., \& {Margalit}, B. 2017, \apj, 841, 14

\bibitem[{{Metzger} {et~al.}(2011){Metzger}, {Giannios}, {Thompson},
  {Bucciantini}, \& {Quataert}}]{Metzger11}
{Metzger}, B.~D., {Giannios}, D., {Thompson}, T.~A., {Bucciantini}, N., \&
  {Quataert}, E. 2011, MNRAS, 413, 2031

\bibitem[{{Metzger} {et~al.}(2008){Metzger}, {Quataert}, \&
  {Thompson}}]{Metzger08}
{Metzger}, B.~D., {Quataert}, E., \& {Thompson}, T.~A. 2008, \mnras, 385, 1455

\bibitem[{{Montanari} {et~al.}(2005){Montanari}, {Frontera}, {Guidorzi}, \&
  {Rapisarda}}]{Montanari05}
{Montanari}, E., {Frontera}, F., {Guidorzi}, C., \& {Rapisarda}, M. 2005, \apj,
  625, L17

\bibitem[{{Nakar} \& {Sari}(2012)}]{Nakar12}
{Nakar}, E., \& {Sari}, R. 2012, ApJ, 747, 88

\bibitem[{{Nicholl} {et~al.}(2017){Nicholl}, {Williams}, {Berger}, {Villar},
  {Alexander}, {Eftekhari}, \& {Metzger}}]{Nicholl17}
{Nicholl}, M., {Williams}, P.~K.~G., {Berger}, E., {et~al.} 2017, \apj, 843, 84

\bibitem[{{Niino}(2018)}]{Niino18}
{Niino}, Y. 2018, \apj, 858, 4

\bibitem[{{Norris} \& {Bonnell}(2006)}]{Norris06}
{Norris}, J.~P., \& {Bonnell}, J.~T. 2006, ApJ, 643, 266

\bibitem[{{Palaniswamy} {et~al.}(2018){Palaniswamy}, {Li}, \&
  {Zhang}}]{Palaniswamy18}
{Palaniswamy}, D., {Li}, Y., \& {Zhang}, B. 2018, \apj, 854, L12

\bibitem[{{Palaniswamy} {et~al.}(2014){Palaniswamy}, {Wayth}, {Trott},
  {McCallum}, {Tingay}, \& {Reynolds}}]{Palaniswamy14}
{Palaniswamy}, D., {Wayth}, R.~B., {Trott}, C.~M., {et~al.} 2014, \apj, 790, 63

\bibitem[{{Palmer} {et~al.}(2005){Palmer}, {Barthelmy}, {Gehrels}, {Kippen},
  {Cayton}, {Kouveliotou}, {Eichler}, {Wijers}, {Woods}, {Granot}, {Lyubarsky},
  {Ramirez-Ruiz}, {Barbier}, {Chester}, {Cummings}, {Fenimore}, {Finger},
  {Gaensler}, {Hullinger}, {Krimm}, {Markwardt}, {Nousek}, {Parsons}, {Patel},
  {Sakamoto}, {Sato}, {Suzuki}, \& {Tueller}}]{Palmer05}
{Palmer}, D.~M., {Barthelmy}, S., {Gehrels}, N., {et~al.} 2005, \nat, 434, 1107

\bibitem[{{Petroff} {et~al.}(2019{\natexlab{a}}){Petroff}, {Hessels}, \&
  {Lorimer}}]{Petroff19_rev}
{Petroff}, E., {Hessels}, J. W.~T., \& {Lorimer}, D.~R. 2019{\natexlab{a}}, The
  Astronomy and Astrophysics Review, 27, 4

\bibitem[{{Petroff} {et~al.}(2016){Petroff}, {Barr}, {Jameson}, {Keane},
  {Bailes}, {Kramer}, {Morello}, {Tabbara}, \& {van Straten}}]{Petroff16}
{Petroff}, E., {Barr}, E.~D., {Jameson}, A., {et~al.} 2016, PASA, 33, e045

\bibitem[{{Petroff} {et~al.}(2019{\natexlab{b}}){Petroff}, {Oostrum},
  {Stappers}, {Bailes}, {Barr}, {Bates}, {Bhandari}, {Burgay}, {Burke-Spolaor},
  {Cameron}, {Champion}, {Eatough}, {Flynn}, {Jameson}, {Johnston}, {Keane},
  {Keith}, {Kramer}, {Levin}, {Morello}, {Ng}, {Possenti}, {Ravi}, {van
  Straten}, {Thornton}, \& {Tiburzi}}]{Petroff19}
{Petroff}, E., {Oostrum}, L.~C., {Stappers}, B.~W., {et~al.}
  2019{\natexlab{b}}, \mnras, 482, 3109

\bibitem[{{Planck Collaboration} {et~al.}(2016){Planck Collaboration}, {Ade},
  {Aghanim}, {Arnaud}, {Ashdown}, {Aumont}, {Baccigalupi}, {Banday},
  {Barreiro}, {Bartlett}, \& et~al.}]{cosmoPlanck15}
{Planck Collaboration}, {Ade}, P.~A.~R., {Aghanim}, N., {et~al.} 2016, A\&A,
  594, A13

\bibitem[{Popov {et~al.}(2018)Popov, Postnov, \& Pshirkov}]{Popov18_rev}
Popov, S.~B., Postnov, K.~A., \& Pshirkov, M.~S. 2018, Physics-Uspekhi, 61, 965

\bibitem[{{Rane} \& {Lorimer}(2017)}]{RaneLorimer17_rev}
{Rane}, A., \& {Lorimer}, D. 2017, Journal of Astrophysics and Astronomy, 38,
  55

\bibitem[{{Rane} {et~al.}(2016){Rane}, {Lorimer}, {Bates}, {McMann},
  {McLaughlin}, \& {Rajwade}}]{Rane16}
{Rane}, A., {Lorimer}, D.~R., {Bates}, S.~D., {et~al.} 2016, \mnras, 455, 2207

\bibitem[{{Ravi}(2019)}]{Ravi19}
{Ravi}, V. 2019, \mnras, 482, 1966

\bibitem[{{Ravi} {et~al.}(2019){Ravi}, {Catha}, {D'Addario}, {Djorgovski},
  {Hallinan}, {Hobbs}, {Kocz}, {Kulkarni}, {Shi}, \& {Vedantham}}]{Ravi19b}
{Ravi}, V., {Catha}, M., {D'Addario}, L., {et~al.} 2019, arXiv e-prints,
  arXiv:1907.01542

\bibitem[{{Romero} {et~al.}(2016){Romero}, {del Valle}, \& {Vieyro}}]{Romero16}
{Romero}, G.~E., {del Valle}, M.~V., \& {Vieyro}, F.~L. 2016, \prd, 93, 023001

\bibitem[{{Rowlinson} {et~al.}(2010){Rowlinson}, {O'Brien}, {Tanvir}, {Zhang},
  {Evans}, {Lyons}, {Levan}, {Willingale}, {Page}, {Onal}, {Burrows},
  {Beardmore}, {Ukwatta}, {Berger}, {Hjorth}, {Fruchter}, {Tunnicliffe}, {Fox},
  \& {Cucchiara}}]{Rowlinson10}
{Rowlinson}, A., {O'Brien}, P.~T., {Tanvir}, N.~R., {et~al.} 2010, \mnras, 409,
  531

\bibitem[{{Scholz} {et~al.}(2016){Scholz}, {Spitler}, {Hessels}, {Chatterjee},
  {Cordes}, {Kaspi}, {Wharton}, {Bassa}, {Bogdanov}, {Camilo}, {Crawford},
  {Deneva}, {van Leeuwen}, {Lynch}, {Madsen}, {McLaughlin}, {Mickaliger},
  {Parent}, {Patel}, {Ransom}, {Seymour}, {Stairs}, {Stappers}, \&
  {Tendulkar}}]{Scholz16}
{Scholz}, P., {Spitler}, L.~G., {Hessels}, J.~W.~T., {et~al.} 2016, \apj, 833,
  177

\bibitem[{{Scholz} {et~al.}(2017){Scholz}, {Bogdanov}, {Hessels}, {Lynch},
  {Spitler}, {Bassa}, {Bower}, {Burke-Spolaor}, {Butler}, {Chatterjee},
  {Cordes}, {Gourdji}, {Kaspi}, {Law}, {Marcote}, {McLaughlin}, {Michilli},
  {Paragi}, {Ransom}, {Seymour}, {Tendulkar}, \& {Wharton}}]{Scholz17}
{Scholz}, P., {Bogdanov}, S., {Hessels}, J.~W.~T., {et~al.} 2017, \apj, 846, 80

\bibitem[{{Shannon} \& {Ravi}(2017)}]{ShannonRavi17}
{Shannon}, R.~M., \& {Ravi}, V. 2017, \apjl, 837, L22

\bibitem[{{Shannon} {et~al.}(2018){Shannon}, {Macquart}, {Bannister}, {Ekers},
  {James}, {Os{\l}owski}, {Qiu}, {Sammons}, {Hotan}, {Voronkov}, {Beresford},
  {Brothers}, {Brown}, {Bunton}, {Chippendale}, {Haskins}, {Leach},
  {Marquarding}, {McConnell}, {Pilawa}, {Sadler}, {Troup}, {Tuthill},
  {Whiting}, {Allison}, {Anderson}, {Bell}, {Collier}, {G{\"u}rkan}, {Heald},
  \& {Riseley}}]{Shannon18}
{Shannon}, R.~M., {Macquart}, J.~P., {Bannister}, K.~W., {et~al.} 2018, \nat,
  562, 386

\bibitem[{{Soderberg} {et~al.}(2006){Soderberg}, {Kulkarni}, {Nakar}, {Berger},
  {Cameron}, {Fox}, {Frail}, {Gal-Yam}, {Sari}, {Cenko}, {Kasliwal},
  {Chevalier}, {Piran}, {Price}, {Schmidt}, {Pooley}, {Moon}, {Penprase},
  {Ofek}, {Rau}, {Gehrels}, {Nousek}, {Burrows}, {Persson}, \&
  {McCarthy}}]{Soderberg06d}
{Soderberg}, A.~M., {Kulkarni}, S.~R., {Nakar}, E., {et~al.} 2006, \nat, 442,
  1014

\bibitem[{{Spitler} {et~al.}(2016){Spitler}, {Scholz}, {Hessels}, {Bogdanov},
  {Brazier}, {Camilo}, {Chatterjee}, {Cordes}, {Crawford}, {Deneva}, {Ferdman},
  {Freire}, {Kaspi}, {Lazarus}, {Lynch}, {Madsen}, {McLaughlin}, {Patel},
  {Ransom}, {Seymour}, {Stairs}, {Stappers}, {van Leeuwen}, \&
  {Zhu}}]{Spitler16}
{Spitler}, L.~G., {Scholz}, P., {Hessels}, J.~W.~T., {et~al.} 2016, \nat, 531,
  202

\bibitem[{{Tendulkar} {et~al.}(2016){Tendulkar}, {Kaspi}, \&
  {Patel}}]{Tendulkar16}
{Tendulkar}, S.~P., {Kaspi}, V.~M., \& {Patel}, C. 2016, \apj, 827, 59

\bibitem[{{Tendulkar} {et~al.}(2017){Tendulkar}, {Bassa}, {Cordes}, {Bower},
  {Law}, {Chatterjee}, {Adams}, {Bogdanov}, {Burke-Spolaor}, {Butler},
  {Demorest}, {Hessels}, {Kaspi}, {Lazio}, {Maddox}, {Marcote}, {McLaughlin},
  {Paragi}, {Ransom}, {Scholz}, {Seymour}, {Spitler}, {van Langevelde}, \&
  {Wharton}}]{Tendulkar17}
{Tendulkar}, S.~P., {Bassa}, C.~G., {Cordes}, J.~M., {et~al.} 2017, \apjl, 834,
  L7

\bibitem[{{Thompson}(1994)}]{Thompson94}
{Thompson}, C. 1994, MNRAS, 270, 480

\bibitem[{{Thornton} {et~al.}(2013){Thornton}, {Stappers}, {Bailes},
  {Barsdell}, {Bates}, {Bhat}, {Burgay}, {Burke-Spolaor}, {Champion}, {Coster},
  {D'Amico}, {Jameson}, {Johnston}, {Keith}, {Kramer}, {Levin}, {Milia}, {Ng},
  {Possenti}, \& {van Straten}}]{Thornton13}
{Thornton}, D., {Stappers}, B., {Bailes}, M., {et~al.} 2013, Science, 341, 53

\bibitem[{{Troja} {et~al.}(2017){Troja}, {Lipunov}, {Mundell}, {Butler},
  {Watson}, {Kobayashi}, {Cenko}, {Marshall}, {Ricci}, {Fruchter}, {Wieringa},
  {Gorbovskoy}, {Kornilov}, {Kutyrev}, {Lee}, {Toy}, {Tyurina}, {Budnev},
  {Buckley}, {Gonz{\'a}lez}, {Gress}, {Horesh}, {Panasyuk}, {Prochaska},
  {Ramirez-Ruiz}, {Rebolo Lopez}, {Richer}, {Roman-Zuniga}, {Serra-Ricart},
  {Yurkov}, \& {Gehrels}}]{Troja17nat}
{Troja}, E., {Lipunov}, V.~M., {Mundell}, C.~G., {et~al.} 2017, Nature, 547,
  425

\bibitem[{{Tsvetkova} {et~al.}(2017){Tsvetkova}, {Frederiks}, {Golenetskii},
  {Lysenko}, {Oleynik}, {Pal'shin}, {Svinkin}, {Ulanov}, {Cline}, {Hurley}, \&
  {Aptekar}}]{KWGRBcat17}
{Tsvetkova}, A., {Frederiks}, D., {Golenetskii}, S., {et~al.} 2017, \apj, 850,
  161

\bibitem[{{Usov}(1992)}]{Usov92}
{Usov}, V.~V. 1992, Nature, 357, 472

\bibitem[{{Virgili} {et~al.}(2009){Virgili}, {Liang}, \& {Zhang}}]{Virgili09}
{Virgili}, F.~J., {Liang}, E.-W., \& {Zhang}, B. 2009, \mnras, 392, 91

\bibitem[{{Wanderman} \& {Piran}(2010)}]{WandermanPiran10}
{Wanderman}, D., \& {Piran}, T. 2010, MNRAS, 406, 1944

\bibitem[{{Wanderman} \& {Piran}(2015)}]{WandermanPiran15}
---. 2015, MNRAS, 448, 3026

\bibitem[{{Waxman} {et~al.}(2007){Waxman}, {M{\'e}sz{\'a}ros}, \&
  {Campana}}]{Waxman07}
{Waxman}, E., {M{\'e}sz{\'a}ros}, P., \& {Campana}, S. 2007, \apj, 667, 351

\bibitem[{{Xi} {et~al.}(2017){Xi}, {Tam}, {Peng}, \& {Wang}}]{Xi17}
{Xi}, S.-Q., {Tam}, P.-H.~T., {Peng}, F.-K., \& {Wang}, X.-Y. 2017, \apjl, 842,
  L8

\bibitem[{{Xu} \& {Han}(2015)}]{XuHan15}
{Xu}, J., \& {Han}, J.~L. 2015, Research in Astronomy and Astrophysics, 15,
  1629

\bibitem[{{Yamasaki} {et~al.}(2016){Yamasaki}, {Totani}, \&
  {Kawanaka}}]{Yamasaki16}
{Yamasaki}, S., {Totani}, T., \& {Kawanaka}, N. 2016, \mnras, 460, 2875

\bibitem[{{Yang} {et~al.}(2019){Yang}, {Zhang}, \& {Zhang}}]{Yang19}
{Yang}, Y.-H., {Zhang}, B.-B., \& {Zhang}, B. 2019, \apjl, 875, L19

\bibitem[{{Zhang}(2018)}]{Zhang18a}
{Zhang}, B. 2018, \apjl, 867, L21

\bibitem[{{Zhang} \& {Zhang}(2017)}]{ZhangZhang17}
{Zhang}, B.-B., \& {Zhang}, B. 2017, \apjl, 843, L13

\bibitem[{{Zhang} {et~al.}(2019){Zhang}, {Hobbs}, {Dai}, {Toomey},
  {Staveley-Smith}, {Russell}, \& {Wu}}]{Zhang19a}
{Zhang}, S.-B., {Hobbs}, G., {Dai}, S., {et~al.} 2019, \mnras, 484, L147

\end{thebibliography}

\appendix

\section{Background modeling}
\label{sec:bkg}
We conveniently split the data set into individual orbits: the data available for each orbit consist of 1-s ratemeters continuously acquired in both 40--700 and $>$100~keV energy bands, covering the time interval from the end of a given passage over the South Atlantic Anomaly to the beginning of the next one. Hereafter, times are referred to the FRB time. 
The FRB orbit, spanning from $-2200$ to $+2770$~s, is affected by no classified transient event, such as GRBs \citep{Frontera09}, solar X--ray flares, or magnetar outburts. We estimated the background rates for each unit and energy band using the data of the fifteenth previous orbit (hereafter, referred to as the ``background orbit''), during which the spacecraft had the same attitude as for the FRB orbit. The central time of the background orbit precedes that of the FRB one by 85505~s, i.e. only $655$~s difference from a sidereal day: this way, not only do both orbits experience the same configuration with respect to the Earth magnetic field, but they also share a very similar visible portion of the sky. Data were cleaned from a few spikes due to high-energy charged particles: this was done by replacing the $>7\sigma$ excesses with respect to a moving average with statistical realizations of the local average, after verifying that the same excesses were not present in different detectors, so as to exclude an electromagnetic wave origin. We therefore interpolated the rates of the background orbit in terms of Legendre polynomials up to degree 20, upon renormalizing the time span of the orbit to the $[-1,1]$ interval. The reason behind this choice lies in their orthogonality and easier comparison between the variance of a given time series, as described by its Legendre spectrum, and that of another series, with respect to using generic polynomials. A more detailed justification will be supplied in a future dedicated work. The quality of the interpolation was verified by applying a normality test on the residuals\footnote{We used the python function ``normaltest'', based on D'Agostino and Pearson's test, as implemented in the {\sc scipy.stats} package.}, a $\chi^2$ test, and a runs test\footnote{We used the python function ``wald\_wolfowitz'' from the {\sc skidmarks} package.} to ensure that no trend was present. All the corresponding p--values were above the 1\% threshold.

We then subtracted the background model for each unit and energy band from the corresponding data of the FRB orbit and checked the quality of the result through the same statistics as above: the worst reduced $\chi^2$ was $1.12$, while the lowest p--value for the runs test was 5\%, thus confirming the robustness of the background modeling. This procedure benefited from the low variability of the equatorial orbit of \sax\,\citep{Frontera97}.

\end{document}